\documentclass{aa}
\usepackage{graphicx}
\usepackage{txfonts}

\newcommand{\revmp}{Rev.~Mod.~Phys.}
\newcommand\sovastl{Soviet~Ast.~Lett.}
\newcommand\E[1]{\times10^{#1}}

\newcommand\U[1]{{\,\rm #1}}
\newcommand\gm{\gamma}
\newcommand\vkp{\varkappa}
\newcommand\om{\omega}
\newcommand\eps{\epsilon}
\newcommand\Rtil{{\tilde R}} \newcommand\ttil{{\tilde t}}
\newcommand\Vtil{{\tilde V}} \newcommand\Etil{{\tilde E}}
\newcommand\dR{\dot R}
\newcommand\ddR{\ddot R}
\newcommand\tendsto{\rightarrow}
\newcommand\rs[1]{_\mathrm{#1}}                 
\newcommand\Rz{R\rs{o}}
\newcommand\tz{t\rs{o}}
\newcommand\dRz{\dot R\rs{o}}
\newcommand\Pz{P\rs{o}}
\newcommand\vkpz{\vkp\rs{o}}
\newcommand\vkpas{\vkp\rs{as}}
\newcommand\vkpsed{\vkp\rs{Sed}}
\newcommand\tauz{\tau\rs{o}}
\newcommand\rz{r\rs{o}}
\newcommand\mz{m\rs{o}}
\newcommand\rhoa{\rho\rs{a}}
\newcommand\na{n\rs{a}}
\newcommand\Ekin{E\rs{kin}}
\newcommand\Eth{E\rs{th}}
\newcommand\Etot{E\rs{tot}}
\newcommand\ttran{t\rs{tran}}
\newcommand\Rtran{R\rs{tran}}
\newcommand\rmax{r\rs{max}}
\newcommand\tmax{t\rs{max}}
\newcommand\taumax{\tau\rs{max}}
\newcommand\mmax{m\rs{max}}
\newcommand\tinters{t\rs{inters}}
\newcommand\Rinters{R\rs{inters}}
\newcommand\tauinters{\tau\rs{inters}}
\newcommand\rinters{r\rs{inters}}
\newcommand\tPDS{t\rs{PDS}}

\begin{document}

\title{Analytic Solutions for the Evolution\\
 of Radiative Supernova Remnants}
\author{R. Bandiera \inst{1} \and O. Petruk \inst{2}}
\offprints{R. Bandiera}
\institute{
  Osservatorio Astrofisico di Arcetri, Largo E.Fermi 5 Firenze 50125, Italy\\
  \email{bandiera@arcetri.astro.it}
\and
  Osservatorio Astrofisico di Arcetri, Largo E.Fermi 5 Firenze 50125, Italy\\
  Institute for Applied Problems in Mechanics and Mathematics, Naukova St. 3-b
  Lviv 79000, Ukraine\\
  \email{petruk@arcetri.astro.it}
}
\date{Received 7 January 2004 / Accepted 10 February 2004}
                                                                                
\abstract{
We present the general analytic solution for the evolution of radiative
supernova remnants in a uniform interstellar medium, under thin-shell
approximation.
This approximation is shown to be very accurate approach to this task.
For a given set of parameters, our solution closely matches the results of
numerical models, showing a transient in which the deceleration parameter
reaches a maximum value of 0.33, followed by a slow convergence to the
asymptotic value 2/7.
Oort (1951) and McKee and Ostriker (1977) analytic solutions are discussed,
as special cases of the general solution we have found.
}

\maketitle

\section{Introduction}

In recent years numerical modelling of the structure and evolution of
supernova remnants (SNRs) has reached an unprecedented level of detail.
Nonetheless analytic models still play a very important role, when general
properties have to be investigated, as well as when direct relations have
to be drawn between pure observational quantities (like size, flux, etc.) and
intrinsic physical parameters.

The adiabatic phase of SNR evolution (in a uniform and homogeneous medium)
is well described by the Sedov (\cite{Sedov-59}) analytic solution, which
reproduces both the SNR radial evolution and its inner structure.
This exact solution has been made possible by the fact that during this
phase the SNR evolution is self-similar.
This is no longer the case when radiative losses become important, and
therefore no exact analytic solution is known for the late SNR evolution.

Approximated solutions in the adiabatic regime and beyond may be also
obtained using a ``thin-shell'' model (see e.g.\ Zel'dovich \& Raizer
\cite{Zel-Raizer-66}, Ostriker \& McKee \cite{Ostr-McKee-88}).
This approach assumes that the whole mass (and therefore kinetic energy)
of the SNR is located in a rather thin shell just behind the outer shock;
while the inner region is filled with a very hot and rarefied gas, of
negligible total mass, but containing most of the SNR internal energy.

For the adiabatic phase this approximation is only moderately accurate
(see e.g.\ Zel'dovich \& Raizer \cite{Zel-Raizer-66}).
In fact, according to the Sedov solution the gas density vanishes in the
inner regions while its pressure keeps finite; however, the outer layer
containing most of the mass is geometrically rather thick.
On the other hand, numerical works (e.g.\ Falle \cite{Falle-75}, Blondin et
al.\ \cite{Blondin-et-al-98}, hereafter BWBR) trace the formation of a much
thinner shell in the radiative phase, therefore indicating that a thin-shell
approximation should be far more accurate in describing the late evolution.

Oort (\cite{Oort-51}) presented a first thin-shell approach to a radiative
SNR expansion.
By assuming momentum conservation in the shell, he found the SNR radius to
evolve as $R\propto t^{1/4}$.
This solution, also known as ``momentum-conserving snowplow'', assumes that
cooling is extremely efficient everywhere (and therefore that the interior
pressure vanishes).
However, numerical models (e.g.\ Chevalier \cite{Chev-74}) show that, even
in the radiative phase, the gas in the central regions becomes so rarefied
that its cooling time still keeps considerably longer than the SNR age.
This led McKee \& Ostriker (\cite{McKee-Ostr-77}) to introduce a
``pressure-driven snowplow'' model, in which a fossil pressure in the hot
interior has a substantial dynamical effect on the outer shell: in this case
the radial evolution is $R\propto t^{2/7}$ (for adiabatic index $\gm=5/3$).

Even though the ``pressure-driven snowplow'' formula gets closer than the
``momentum-conserving snowplow'' one to the numerical results, some discrepancy
still remains.
For instance, by defining the ``deceleration parameter'' as $m=d\log R/d\log
t$, numerical models obtain an asymptotic value ranging from 0.31 (Chevalier
\cite{Chev-74}) to 0.33 (BWBR).
These values are significantly different from the analytic value, $2/7$ (namely
0.286), and various authors have discussed the origin of such discrepancy.
Cioffi et al.\ (\cite{Cioffi-et-al-88}) ascribe it to a ``memory'' of the
previous Sedov phase, leading to an actual internal pressure larger than
that derived from the analytic model.
BWBR, instead, attribute this discrepancy to the influence of the reverse
shock, which moves towards the center raising the thermal energy, thus
leading to a milder deceleration.

Other authors have estimated analytically the radial evolution under more
general conditions than those given above.

Ostriker \& McKee (\cite{Ostr-McKee-88}) have shown that, for a general $\gm$
as well as a power-law ambient density profile ($\rhoa(r)\propto r^{-\om}$),
$m=1/(4-\om)$ for a ``momentum-conserving snowplow'', while $m=2/(2+3\gm-\om)$
for a ``pressure-driven snowplow''.

Liang \& Keilty (\cite{Liang-Keilty-2000}) have considered the case in which
only a (constant) fraction $\eps$ of the kinetic energy of the incoming flow
is radiated in the outer shock.
For $\gm=5/3$, $m$ is found to decrease quasi-linearly with $\eps$, from
$2/5$ for the adiabatic case ($\eps=0$) to $2/7$ for the fully radiative case
($\eps=1$); and a value of $\eps$ of about 0.8 (0.6) is required in order
to obtain $m=0.31$ (0.33), as indicated by the numerical models.
However, while $\eps<1$ may be appropriate to describe gamma-ray burst
afterglows (Cohen et al.\ \cite{Cohen-98}), SNR radiative shocks should be
described as fully radiative ones (namely with $\eps$ very close to unity).

The effect of cooling in the hot interior on the deceleration parameter has
been studied by Gaffet (\cite{Gaffet-83}), with the following results.
Adiabaticity holds throughout most of the volume occupied by the hot gas,
while cooling occurs only near the boundary with the radiative shell, giving
as effect a net mass transfer from the hot interior to the shell.
Assuming that the gas in the hot interior follows a cooling law $\Lambda\propto
T^{-c}$, this paper discusses different regimes for different choices of $\gm$
and $c$, showing that the asymptotic value of $m$ must be in the range between
the values $1/4$ (Oort limit) and $2/(2+3\gm)$ (McKee and Ostriker limit).

When $c>(2/3)(5\gm-8)/(3-2\gm)$ ($c>-2/3$ for $\gm=5/3$) the asymptotic
value of $m$ is 1/4.
This is the case for cooling functions typical of the SNR regime, where 
$c>0$ and is usually taken in the range from 0.5 (e.g.\ Cioffi et al.\
\cite{Cioffi-et-al-88}) to 1.0 (BWBR).
Therefore, according to this result, after the commonly known radiative
phase there could be a very late evolutive phase, in which 
radiative losses of hot interior begin to be prominent and the SNR evolves as
``momentum-conserving snowplow''.
However, as it may be derived from numerical results, in a typical SNR, 
radiative cooling of the hot interior is negligible until very late times.
Therefore the onset of the ``momentum-conserving snowplow'' regime should
occur only near the end of a SNR lifetime, or not to occur at all.

A common limitation of all above-mentioned analytical models is that the
radial evolution of the radiative shock has been approximated by a power-law
behaviour $R\propto t^{m}$ (with constant $m$).
This allows a simplified treatment of radiative SNR evolution; however, it
is natural to expect that a power-law expansion occurs only at late times
(i.e.\ at large $R$ values), after the transition from adiabatic to radiative
expansion has been completed.

In this paper we shall show:
1) that the quoted difference between the numerical and asymptotic analytic
value is just a consequence of the fact that the time needed to reach the
asymptotic power-law regime is long compared with the age of the SNR;
2) that the SNR radial evolution during the transient phase is adequately
described by a thin-shell model;
3) that a general analytic solution of this problem exists. 

The plan of the paper is as follows.
In Sect.~2 we derive a differential equation that describes the radiative SNR
evolution, find its complete solution for an arbitrary adiabatic coefficient,
discuss the general properties for the two branches of solutions that we find,
and show that the Oort solution is just a special case of our general solution.
Sect.~3 focuses on the conventional case $\gm=5/3$, which allows simpler
analytic formulae, and shows that our solution tends to the asymtotic regime
given by the solution of McKee \& Ostriker (\cite{McKee-Ostr-77}); with
the use of numerical results, we also derive the most appropriate initial
conditions for our solution.
Sect.~4 concludes.

\section{Equations and solutions for a general adiabatic index}

Let us consider a fully radiative shock expanding into a uniform medium and
neglect the cooling of the hot interior.
In the thin-shell approximation, mass, momentum, and central pressure
evolution are described by the following set of equations:
\begin{eqnarray}
\frac{dM}{dt}\quad\,&=&4\pi\rhoa R^2\dR, \label{masscons}\\
\frac{d(M\dR)}{dt}&=&4\pi PR^2, \label{momentumcons} \\
\frac{dP}{dt}\quad\,\;&=&-3\gm P\frac{\dR}{R}, \label{innerpcons}
\end{eqnarray}
where $R$ and $\dR$ are respectively shock radius and velocity, $M$ is the
mass of the radiative shell, $P$ is the pressure of the (adiabatically
evolving) inner region, $\gm$ is the adiabatic coefficient, and $\rhoa$
is the (constant) density of the ambient medium.

The above equations can be reduced to a single one:
\begin{equation}
  \ddR+\frac{3\dR^2}{R}=\frac{3\Pz\Rz^{3\gm}}{\rhoa}R^{-3\gm-1},
 \label{singleeq}
\end{equation}
where the quantities $\Rz$ and $\Pz$ indicate respectively the SNR radius
and the pressure of the hot cavity at a reference time ($\tz$), that can be
arbitrarily chosen.

In order to solve analytically Eq.~(\ref{singleeq}) for a general value of
$\gm$ (with the condition $1<\gm<2$) let us first define the quantity:
\begin{equation}
  K=\frac{2}{2-\gm}\frac{PR^{3\gm}}{\rhoa}.
\end{equation}
$K$ is constant in time, and therefore it can be evaluated in terms of
quantities at the time $\tz$.
By using the substitution $w(R)=\dR^2$, Eq.~(\ref{singleeq}) trasforms into:
\begin{equation}
  \frac{dw}{dR}+6\frac{w}{R}=3(2-\gm)KR^{-3\gm-1},
\end{equation}
that is a linear differential equation and can then be easily integrated.
Its general solution is:
\begin{equation}
  w=K(R^{-3\gm}-HR^{-6}),
 \label{sollin}
\end{equation}
where the constant $H$ is, at any time, equal to:
\begin{equation}
  H=R^{3(2-\gm)}\left(1-\frac{(2-\gm)\rhoa}{2P}\dR^2\right).
  \label{hdef}
\end{equation}
In particular, it may be expressed in terms of quantities at the time $\tz$.
Depending on the sign of $H$, there are two different branches of solutions.
By evaluating the kinetic energy of the shell and the thermal energy of the
inner hot bubble respectively as:
\begin{eqnarray}
  \Ekin&=&\frac{4\pi R^3}{3}\frac{\rhoa\dR^2}{2},	\label{ekin}\\
  \Eth &=&\frac{4\pi R^3}{3}\frac{P}{\gm-1}, \label{eth}
\end{eqnarray}
Eq.~(\ref{hdef}) shows that, in the two branches, the energy ratio
$\vkp=\Ekin/\Eth=(\gm-1)\rhoa\dR^2/2P$ is respectively less ($H$-positive
case) and greater ($H$-negative case) than $(\gm-1)/(2-\gm)$, and that time
evolution does not change the sign of this inequality.

Let us label these two branches  of solutions as ``slow'' and ``fast'',
depending whether the kinetic energy is respectively less ($H$ positive case)
or greater ($H$ negative case) than $(\gm-1) \Eth/(2-\gm)$; or, equivalently,
less or greater than $(\gm-1)\Etot$ (where $\Etot=\Ekin+\Eth$).
The choice of the appropriate branch of solutions only depends on the
initial conditions.

Although in the next section we shall see that the slow case is that physically
relevant for the SNR evolution, let us discuss here both branches.
When double signs are shown, in some of the following equations, the convention
used is that the upper sign refers to the slow branch, while the lower sign
to the fast branch.
Once defined space and time scale units as
\begin{eqnarray}
  \Rtil&=&(\pm H)^{1/3(2-\gm)};
  \label{rtdef}\\
  \ttil&=&(\pm H)^{(2+3\gm)/6(2-\gm)}K^{-1/2},
  \label{ttdef}
\end{eqnarray}
and introduced the dimensionless space and time coordinates $r=R/\Rtil$,
$\tau=t/\ttil$, the evolution in size follows the equation:
\begin{equation}
  \frac{dr}{d\tau}=\sqrt{r^{-3\gm}\mp r^{-6}},
  \label{dimless}
\end{equation}
that can be integrated to give $\tau(r)$.
The dimensional velocity is obtained multiplying $dr/d\tau$ by the velocity
scale $\Vtil=\Rtil/\ttil$.
It is evident that, while fast solutions extend to all positive values of $r$,
solutions in the slow branch are real only for $r\ge1$.

For a general value of $\gm$, the solution involves hypergeometric functions
($F$), and can be written as:
\begin{eqnarray}
  \tau\rs{F}(r)&=&\frac{r^4}{4}F\left(\frac{1}{2},\frac{4}{3(2-\gm)},
    1+\frac{4}{3(2-\gm)};-r^{3(2-\gm)}\right)	\nonumber\\
  &&-\frac{1}{4}F\left(\frac{1}{2},\frac{4}{3(2-\gm)},
    1+\frac{4}{3(2-\gm)};-1\right)+C
  \label{fgensol}
\end{eqnarray}
for the fast branch, and as:
\begin{eqnarray}
  \tau\rs{S}(r)&=&\frac{ir^4}{4}F\left(\frac{1}{2},\frac{4}{3(2-\gm)},
    1+\frac{4}{3(2-\gm)};r^{3(2-\gm)}\right)	\nonumber\\
  &&-\frac{i}{4}F\left(\frac{1}{2},\frac{4}{3(2-\gm)},
    1+\frac{4}{3(2-\gm)};1\right)+C
  \label{sgensol}
\end{eqnarray}
for the slow branch.
The time evolution of the SNR radius is obtained by inverting the above
equations.
Note that individual terms in Eq.~(\ref{sgensol}) are complex, but when
$r\ge1$ their imaginary parts cancel out.
Eqs.~(\ref{fgensol}) and (\ref{sgensol}) contain an arbitrary constant, $C$;
both equations have been written here in such a way that $C=\tau(1)$.

A quantity useful to describe the evolution is the deceleration parameter $m$.
The general formula for this quantity is rather complex, but its asymptotic
behaviour at large values of $r$ may be evaluated as:
\begin{equation}
  m(r)=\frac{2}{2+3\gm}\pm\frac{6(2-\gm)}{(9\gm-10)(2+3\gm)}r^{-3(2-\gm)}
    +{\cal O}\left(r^{-6(2-\gm)}\right).
  \label{masymp}
\end{equation}
This power expansion is valid in the range $10/9<\gm<2$.
Note that the limit $2/(2+3\gm)$ is the same found by Ostriker and McKee
(\cite{Ostr-McKee-88}).

In terms of the dimensionless quantity $r$, the ratio of kinetic and thermal
energies is:
\begin{equation}
  \vkp=\frac{\gm-1}{2-\gm}\left(1\mp r^{-3(2-\gm)}\right).
\end{equation}
Therefore at late times the asymptotic value of this ratio is
$\vkpas=(\gm-1)/(2-\gm)$ for both branches.

The ratio between internal and shock pressure is:
\begin{equation}
  \frac{P}{P\rs{s}}=\frac{(\gm-1)(\gm+1)}{4\vkp}=
    \frac{(2-\gm)(\gm+1)}{4(1\mp r^{-3(2-\gm)})},
\end{equation}
then leading to the asymptotic value $(2-\gm)(\gm+1)/4$.

The total energy follows the evolutive law:
\begin{equation}
  \Etot=\Etil r^3\left(\frac{r^{-3\gm}}{\gm-1}\mp r^{-6}\right).
\end{equation}
where $\Etil=2\pi\rhoa\Vtil^2\Rtil^3/3$.
It is easy to show that, in the range of validity of the solutions, $d\Etot/dr$
is always negative, as expected for a radiative solution.

Finally, it can be shown that the original Oort (\cite{Oort-51}) solution,
$R\propto t^{1/4}$, is just a special case of the fast-branch solution.
Neglecting the pressure effect leads the right side of Eq.~(\ref{singleeq})
to vanish.
Since $P=0$ implies the energy ratio $\vkp$ to diverge (and therefore to be
larger than $(\gm-1)/(2-\gm)$), the solution must belong to the fast branch.
From Eq.~(\ref{sollin}) it is apparent that $K$ is required to vanish, while
$H\tendsto-\infty$, in such a way that the product $-KH$ be equal to $R^6\dR^2$
(being this a constant, it can be then evaluated in terms of $\Rz$ and $\dRz$).
Using Eqs.~(\ref{rtdef}) and (\ref{ttdef}), it can be shown that both $\Rtil$
and $\ttil$ diverge, so that this solution must be limited to vanishingly
small $r$ and $\tau$ values.
Therefore Eq.~(\ref{dimless}) simplifies into $dr/d\tau=r^{-3}$, which admits
the solution $\tau(r)=r^4/4-1/4+C$, where we have defined $C=\tau(1)$ for
consistency with the formulation given in Eq.~(\ref{fgensol}).
For $C=1/4$, which means $\tau(0)=0$, we simply have $\tau(r)=r^4/4$ that,
when inverted, gives the Oort's law $r\propto t^{1/4}$.
The same result can be extracted from the general solution,
Eq.~(\ref{fgensol}), by using the fact that $F(a,b,c;x)\tendsto1$ when
$x\tendsto0$.

\section{The slow branch of solutions for $\gm=5/3$}

In the standard case $\gm=5/3$, Eqs.~(\ref{fgensol}) and (\ref{sgensol})
get a much simpler functional dependence, respectively:
\begin{eqnarray}
  \tau\rs{F}(r)&=&\frac{2}{35}\sqrt{r+1}(5r^3-6r^2+8r-16)+\frac{18\sqrt{2}}{35}
    +C,
  \label{fastsol}\\
  \tau\rs{S}(r)&=&\frac{2}{35}\sqrt{r-1}(5r^3+6r^2+8r+16)+C,
  \label{slowsol}
\end{eqnarray}
where again we use $C=\tau(1)$.

The deceleration parameter $m$, for the two branches, evaluates:
\begin{eqnarray}
  m\rs{F}(r)&=&\frac{2}{35}\frac{r+1}{r^4}(5r^3-6r^2+8r-16)+\nonumber\\
  &&\hbox to3cm{\hfil}
    \frac{\sqrt{r+1}}{r^4}\left(C+\frac{18}{35}\sqrt{2}\right),
    \label{mfast}\\
  m\rs{S}(r)&=&\frac{2}{35}\frac{r-1}{r^4}(5r^3+6r^2+8r+16)+
    \frac{\sqrt{r-1}}{r^4}C.
    \label{mslow}
\end{eqnarray}
The asymptotic behaviour at large values of $r$ is:
\begin{equation}
  m(r)=\frac{2}{7}\pm\frac{2}{35r}+{\cal O}\left(\frac{1}{r^2}\right)
\end{equation}
(which is consistent with the more general Eq.~(\ref{masymp})).
Therefore, for $r$ approaching to infinity, in both branches $m$ tends to the
value $2/7$, namely to the asymptotic solution given by McKee and Ostriker
(\cite{McKee-Ostr-77}).
However, an analysis of Eqs.~(\ref{mfast}) and (\ref{mslow}) show different
properties for the two branches.
In particular, only in the slow branch $m(r)$ shows a local maximum.
In a given solution, the position of the maximum and the value reached by $m$
are related by:
\begin{equation}
  \mmax=\frac{2(\rmax-1)}{7\rmax-8},
\end{equation}
valid for $\rmax>8/7$.
Therefore, $\mmax$ is always larger than $2/7$ for any solution in the slow
branch, and it can be considerably larger than $2/7$, if $\rmax$ gets close
to $8/7$.
Furthermore, if during its evolution $m(r)$ is larger than $2/7$ and still
increasing with $r$, it must reach a maximum before approaching the asymptotic
value $2/7$, and then it must belong to the slow branch.
This is what shown by BWBR (their Fig.~3, actually limited to the increasing
part): therefore in the following we shall consider only the slow branch
of solutions.

\begin{figure}
   \centering
   \includegraphics[bb=18 144 568 538,width=8.8cm,clip]{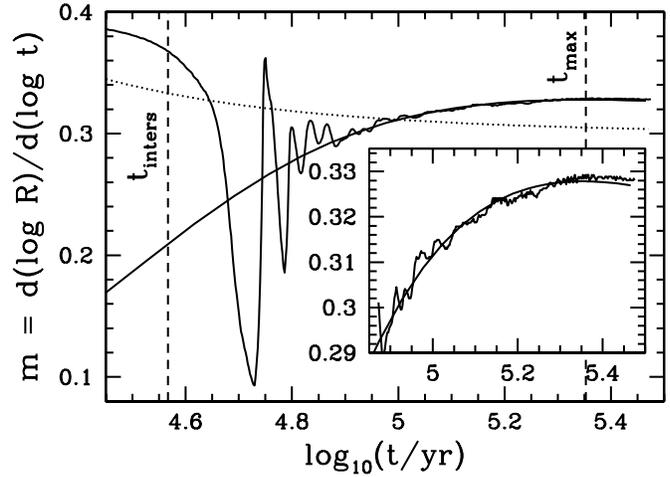}
\caption{
Our best fit solution (slow branch, $C=-0.248$, $\ttil=3.64\E{4}\U{yr}$)
compared with numerical data from BWBR.
The small frame shows just the data used for the fit.
With the exception of strong oscillations in the early transient, the analytic
solution closely describes also the evolution at earlier times.
Dashed lines indicate the positions of $\tmax$ and $\tinters$, while
the dotted line is obtained by using the analytic fit by Cioffi et al.\
\cite{Cioffi-et-al-88} (see text).
}
         \label{fig_fit}
   \end{figure}

Let us use the numerical results from BWBR to determine the most appropriate
parameters for our analytic solution.
For the fit we use the evolution of $m$ (BWBR, their Fig.~3), excluding the
oscillatory transient: the numerical data fitted are for times ranging from
$7.4\E{4}$ till $3.0\E{5}\U{yr}$.
The best (least-square) fit is obtained for $C=-0.248\pm0.006$; while the time
scale inferred from this fit allows us to fix $\ttil=(3.64\pm0.05)\E{4}\U{yr}$
(for all best fit quantities, here we also indicate their 1-$\sigma$ error).
In Fig.~\ref{fig_fit}, the best fit curve is shown against the numerical data.
The best fit curve reaches its maximum value ($\mmax=0.328$) at $\rmax=2.11$
(i.e.\ at $\taumax=6.18$).
Moreover, using Fig.~8 (velocity evolution) from the BWBR, we derive
$\Rtil=17.6\pm0.1\U{pc}$.
From these quantities, the dimensional scaling for energy is
$\Etil=(1.51\pm0.07)\E{51}\U{erg}$.

The numerical simulation we refer to corresponds to the following basic
physical parameters: energy of the explosion, $E_{51}=1$ (in units of
$10^{51}\U{erg}$); and hydrogen ambient density, $\na=0.84\U{cm^{-3}}$.
Adopting the same definitions of BWBR for the transition time, and
corresponding SNR radius
\begin{eqnarray}
  \ttran&\approx&2.9\E{4}E_{51}^{4/17}\na^{-9/17}\U{yr},
    \label{ttran}\\
  \Rtran&\approx&19.1E_{51}^{5/17}\na^{-7/17}\U{pc},
    \label{rtran}
\end{eqnarray}
we determine the dimensionless quantities $\ttil/\ttran=1.14$ and
$\Rtil/\Rtran=0.85$.
All these quantities, although obtained after the comparison with a specific
numerical simulation, can be taken of general validity, for a SNR expanding
in a homogeneous medium, because the analytic solutions allow scaling.

Fig.~1 shows also (dotted curve) the analytic fit as derived from Cioffi et
al.\ (\cite{Cioffi-et-al-88}).
Cioffi et al.\ (\cite{Cioffi-et-al-88}) use a different cooling function
from that of BWBR, whose simulation is shown in Fig.~1.
Therefore in Cioffi et al.\ (\cite{Cioffi-et-al-88}) the functional dependence
of time and length scales on the model parameters are different from those
given in Eqs.~(\ref{ttran}) and (\ref{rtran}).
In order to compare our results with Cioffi et al.\ (\cite{Cioffi-et-al-88})
analytic fit, we have evaluated the scaling time $\ttran$ (labelled as $\tPDS$
in their Eqs.~(3.10) and (3.11)), using $E_{51}=1$ and $\na=0.84\U{cm^{-3}}$
(with solar abundances), obtaining $\ttran=14670\U{yr}$.
The formula used for the deceleration parameter is $m=0.3/(1-\ttran/4t)$,
as derived from Eqs.~(3.32) and (3.33) in that paper.
It is apparent, from Fig.~1, that Cioffi et al.\ (\cite{Cioffi-et-al-88})
fit does not trace the evolution of the deceleration parameter $m$.

Since the evolution of the SNR radius is a continuous function of time, let
us compute the time at which the radiative solution intersects the Sedov one.
With the parameters given above, it happens at $\tinters=1.16\ttran$
(when $\Rinters=1.06\Rtran$), namely, using our dimensionless variables,
at $\tauinters=1.01$ (with $\rinters=1.24$).
At this time, the SNR kinetic, Eq.~(\ref{ekin}), and thermal,
Eq.~(\ref{eth}), energies are respectively $\Ekin=0.191\E{51}\U{erg}$
and $\Eth=0.489\E{51}\U{erg}$, equivalent to a total energy
$\Etot=0.680\E{51}\U{erg}$, and to an energy ratio $\vkp=0.390$.

Nicely, although fortuitously, at $\tinters$ the value of $\tau$ is very
close to unity, while that of $\vkp$ is very close to the Sedov value
($\vkpsed=0.394$).
We could then use $\tauz=1$ and $\vkpz=\vkpsed$ as an approximate criterion,
from which to derive, analytically, that:
1) the solution must belong to the slow branch, since $\vkpz<\vkpas$;
2) $\rz$ and $\mz$, evaluated using the relatioships:
\begin{eqnarray}
  \rz&=&2/(2-\vkpz),\\
  \mz&=&\tauz(2-\vkpz)^{7/2}\vkpz^{1/2}/16,
\end{eqnarray}
are respectively $\rz\simeq1.245$ and $\mz\simeq0.206$;
3) $C$, evaluated using Eq.~(\ref{slowsol}), is $\simeq-0.272$ (to be compared
with the best fit value $-0.248$).

\section{Conclusions}

Using a thin-shell approach, we have developed the definitive analytic
treatment for the evolution of a SNR in the radiative phase, and we have
also obtained a series of interesting relations.
The main findings of the present work are the following.

The discrepancy between the analytic prediction of the asymptotic value of
the deceleration parameter ($m=2/7$, McKee and Ostriker \cite{McKee-Ostr-77})
and that derived numerically ($m=0.33$, BWBR) is only apparent.
This discrepancy has been attributed to the presence of a reverse shock
moving towards the center.
We show, instead, that a thin-shell model, that by definition does not
contain any information on inner structure details, closely fits the SNR
evolution as derived numerically.

We then confirm that $2/7$ is the correct asymptotic value, even though the
convergence towards this value is expected to be slow.
We believe that, if BWBR numerical simulation had been runned until later
stages of the SNR evolution, it would have shown that $m$ does not keep
constant to $0.33$, but eventually decreases.
This has been already pointed out by Chevalier (\cite{Chev-74}) and can
be seen in Fig.~5 of Cioffi et al.\ (\cite{Cioffi-et-al-88}), in Fig.~3
of Falle (\cite{Falle-81}), and in Fig.~4b of Mansfield \& Salpeter
(\cite{Mansf-Salp-74}).
However, the convergence to the asymptotic value may need times longer than
the SNR lifetime.

It might be expected that the evolution will eventually change from a
``pressure-driven snowplow'' ($m=2/7$, McKee and Ostriker \cite{McKee-Ostr-77})
to a ``momentum-conserving snowplow'' ($m=1/4$, Oort \cite{Oort-51}), as
a consequence that the right side of Eq.~(\ref{singleeq}) vanishes when
$R\tendsto\infty$.
However, Cioffi et al.\ (\cite{Cioffi-et-al-88}) have noted that, even at
very late times ($\sim10^2\ttran$), the deceleration parameter $m$ is still
closer to $2/7$ than to $1/4$.
We have shown that such evolutive transition may in fact not occur, because
the two kinds of evolution are associated with two different branches of
solutions, corresponding to different initial conditions.
In other words, for a deceleration parameter smaller than $2/7$, the right
side of Eq.~(\ref{singleeq}) (with $\gm=5/3$) vanishes more slowly than the
left side.
Therefore, unless the pressure term is negligible from the beginning, or part
of the internal energy of the hot interior is lost by other channels (e.g.\
by electron conduction, or radiative processes), pressure effects must play
an important role in the evolution at any time, until the SNR merges with
the ambient medium.

Some of the conclusions we have presented here could have been reached
long before.
In fact, Blinnikov et al.\ (\cite{Blinn-et-al-82}) have reduced the system
of equations for the thin-shell evolution to the single equation, equivalent
to our Eq.~(\ref{singleeq}), and have obtained a solution equivalent to
our Eq.~(\ref{slowsol}).
However, they have not discussed the properties of this solution, taking
that it would have quickly relaxed to the asymptotic behaviour.

\acknowledgements{This work has been supported by MIUR under grant
Cofin2001--02--10 and by MIUR under grant Cofin2002.
OP acknowledges grant MIUR Cofin2001--02--10 for his fellowship.}



\end{document}